# Indication of intrinsic room-temperature ferromagnetism in $Ti_{1-x}Co_xO_{2-\delta}$ thin film: An x-ray magnetic circular dichroism study

K. Mamiya and T. Koide

*Photon Factory, IMSS, High Energy Accelerator Research Organization, Tsukuba, Ibaraki 305-0801, Japan*

A. Fujimori

*Department of Complexity Science and Engineering, University of Tokyo, Kashiwa, Chiba 277-8561, Japan*

H. Tokano

*Institute of Applied Physics, University of Tsukuba, Tsukuba, Ibaraki 305-8573, Japan*

H. Manaka

*Department of Nano-Structure and Advanced Materials, Graduate School of Science and Engineering, Kagoshima University, Kagoshima, Kagoshima 890-0065, Japan*

A. Tanaka

*Department of Quantum Matter, Hiroshima University, Higashi-Hiroshima, Hiroshima 739-8530, Japan*

H. Toyosaki and T. Fukumura

*Institute for Materials Research, Tohoku University, Sendai, Miyagi 980-8577, Japan*

M. Kawasaki

*Institute for Materials Research, Tohoku University, Sendai, Miyagi 980-8577, Japan and Combinatorial Materials Exploration and Technology, Tsukuba, Ibaraki 305-0044, Japan*

Soft x-ray magnetic circular dichroism (XMCD) measurements at the Co $L_{2,3}$ edges of Co doped rutile $TiO_2$ at room temperature have revealed clear multiplet features characteristic of ferromagnetic $Co^{2+}$ ions coordinated by $O^{2-}$ ions, being in sharp contrast to the featureless XMCD spectrum of Co metal or metallic clusters. The absorption and XMCD spectra agree well with a full atomic-multiplet calculation for the $Co^{2+}$ high-spin state in the $D_{2h}$-symmetry crystal field at the Ti site in rutile $TiO_2$. The results indicate that the ferromagnetism arises from the $Co^{2+}$ ions substituting the $Ti^{4+}$ ions.

Dilute magnetic semiconductors with room-temperature ferromagnetism are promising materials for novel spintronics devices, in which both spin and charge degrees of freedom are utilized[1,2]. The finding of ferromagnetism at room temperature in Co-doped rutile and anatase $TiO_2$[3,4] has triggered subsequent intensive studies of various oxide semiconductors doped with magnetic ions[5,6]. Shinde *et al.* found on ferromagnetic nanoparticles which vanish with high-temperature annealing and on intrinsic ferromagnetism remaining after the heat treatment in anatase-type $Ti_{1-x}Co_xO_{2-\delta}$ films[7]. Measurements of the anomalous Hall effect (AHE)[8] and visible-UV magnetic circular dichroism (MCD)[9] in rutile-type $Ti_{1-x}Co_xO_{2-\delta}$ indicated that charge carriers are responsible for intrinsic ferromagnetism. However, the claim for the intrinsic ferromagnetism by AHE was disputed by subsequent work[10], suspecting that Co nanoparticles may be responsible for the AHE. Thus, whether the ferromagnetism is intrinsic or extrinsic has been the subject of much debate.

Soft x-ray magnetic circular dichroism (XMCD) in core-level absorption is especially suitable for approaching the present issue, because it provides element-specific, valence-orbital-selective, microscopic information about the electronic and magnetic states. XMCD measurements were reported on, e.g., Mn-doped GaAs[11] and on Co-doped ZnO[12], showing ferromagnetism at low temperatures. Shimizu *et al.*[13] measured Co $K$-edge x-ray absorption spectra (XAS) in anatase-type Co-doped $TiO_2$ films, confirming the valence state of Co as 2+, and detected an XMCD signal, which probes the Co $4p$ states directly, but not the magnetic Co $3d$ states. Kim *et al.*[14] have studied Co-doped anatase-type $Ti_{1-x}Co_xO_2$ thin films with various $x$ by XMCD measurements at the Co $L_{2,3}$ edges. On the contrary, they observed an XMCD spectral line shape nearly identical to that of metallic Co, and found that this XMCD signal increased with annealing the sample. It was therefore suspected that the ferromagnetism in $Ti_{1-x}Co_xO_2$ was due to segregated metallic Co clusters[14]. Thus, it is still an open question whether the reported room-temperature ferromagnetism is intrinsic or extrinsic.

Here, we report on a combined experimental and theoretical Co $L_{2,3}$-edge XMCD study of rutile-type $Ti_{0.97}Co_{0.03}O_{2-\delta}$ as-deposited films. We successfully observed clear multiplet features at the Co $L_{2,3}$ edges in the XMCD spectrum corresponding to those in XAS of $Ti_{0.97}Co_{0.03}O_{2-\delta}$. The experimentally observed XMCD multiplet features agree qualitatively well with the results of a full atomic-multiplet calculation for high-spin $Co^{2+}$ ions under $D_{2h}$ crystal-field symmetry around the Ti site in rutile $TiO_2$. Our experimental and theoretical observations strongly indicate intrinsic ferromagnetism arising from $Co^{2+}$ ions substituting the $Ti^{4+}$ ions.

Rutile-type $Ti_{0.97}Co_{0.03}O_{2-\delta}$ epitaxial films were synthesized by the pulsed laser deposition method at 400 ºC in oxygen pressures of $1 \times 10^{-7}$ Torr. Ferromagnetism was confirmed at room temperature for the present films by both anomalous Hall effect and magnetization measurements[8]. The present samples also showed an MCD signal in the visible-UV region at room temperature[9]. XAS and XMCD spectra at the Co $L_{2,3}$ edges were measured on bending-magnet (BL-11A) and helical-undulator (AR-NE1B) beam lines at the Photon Factory. Polarization-dependent high resolution XAS were measured at room temperature with the total electron-yield method without Ar ion sputtering or annealing in order to avoid the segregation of metallic Co nanoclusters induced by surface treatments[14]. Magnetic fields of ±1 T were applied to the samples.

Figure 1 shows the Co $L_{2,3}$-edge XAS and XMCD spectra of the as-deposited rutile-type $Ti_{0.97}Co_{0.03}O_{2-\delta}$ film. Here, $\mu_+$ and $\mu_-$ stand for the absorption coefficients for the photon helicity, $h$, parallel and antiparallel to the Co $3d$ majority-spin direction, respectively. Both the XAS and XMCD ($\Delta\mu = \mu_+ - \mu_-$) spectra have been corrected for the degree of circular polarization of the incident light. The XAS spectra of the rutile-type $Ti_{0.97}Co_{0.03}O_{2-\delta}$ thin film showed multiplet features. In the following, we refer to each multiplet feature as A-G. The XMCD spectrum showed clear multiplet features almost one-to-one corresponding to those in the XAS.

Figure 2 shows the expanded XAS and XMCD spectra of $Ti_{0.97}Co_{0.03}O_{2-\delta}$ in the Co $L_3$-edge region. The XMCD spectrum of metallic Co is also shown in Fig. 2 for comparison. A striking point here is that the XMCD spectrum of $Ti_{0.97}Co_{0.03}O_{2-\delta}$ showed a clear negative peak at the energy corresponding to feature D in the XAS spectrum. It is emphasized that this negative XMCD peak at D was not observed in the XMCD spectrum of metallic Co. The dominant negative peak in the XMCD spectrum showed a line shape more flattened than that in the XMCD spectrum of metallic Co, indicating overlapping, unresolved multiplet features corresponding to features B and C in the XAS spectrum. The XMCD spectrum showed a positive peak corresponding to feature A in the XAS spectrum. No corresponding feature was seen in the XMCD spectrum of metallic Co. Similarly, the XAS and the XMCD spectra at the Co $L_2$ edge showed corresponding multiplet structures F and G (Fig. 1). The present XMCD spectra of $Ti_{0.97}Co_{0.03}O_{2-\delta}$ are distinctly different from the previous results by Kim *et al.*[14], where they studied $Ti_{1-x}Co_xO_2$ samples heat-treated prior to the measurements. They observed no multiplet features at the Co $L_{2,3}$ edges in their XMCD spectra, attributing the XMCD to segregated metallic Co clusters. The segregated Co clusters in their samples obviously arose from annealing in a vacuum, as indicated by the systematic increase of XAS and XMCD signals of metallic Co[14], and may not be due to the intrinsic properties of $Ti_{1-x}Co_xO_2$. In contrast, the present experiment clearly revealed the multiplet features in the XMCD spectrum corresponding to those in the XAS without any annealing, verifying intrinsic ferromagnetism arising from $Co^{2+}$ ions that substitute $Ti^{4+}$ ions and are coordinated by $O^{2-}$ ions.

We now compare in Fig. 3 the experimental XAS and XMCD spectra with the results of full atomic-multiplet calculations. The method of calculation is described elsewhere[15]. The calculations were made for a low-spin $Co^{2+}$ ion in a crystal field with $O_h$ symmetry, the high-spin $Co^{2+}$ ion in a crystal field with $O_h$ and $D_{2h}$ symmetry. $D_{2h}$ is the local symmetry around the Ti site in the rutile-type structure. The possibility of the low-spin $Co^{2+}$ ion is immediately excluded from a clear disagreement of the experimental XAS and XMCD in line shape with the theoretical ones. The feature corresponding to D at the $L_3$ edge in the XAS is theoretically reproduced for the high-spin $Co^{2+}$ ion under both $O_h$ and $D_{2h}$ symmetries. Although the overall XMCD spectral line shape is similar between $O_h$ and $D_{2h}$ symmetries, the two features F and G at the $L_2$ edge (Fig. 1) in both the theoretical XAS and XMCD, and the intensity ratio of the theoretical XMCD at the $L_2$ edge to that at the $L_3$ edge shows better agreement with experiment for $D_{2h}$ than for $O_h$. The experimental XAS and XMCD spectra thus show qualitatively the best agreement with the calculated spectra for the $Co^{2+}$ high-spin configuration in the $D_{2h}$ crystal field. This result further supports that ferromagnetic Co ions have the $Co^{2+}$ high-spin electron configuration, and are located at the Ti sites of the rutile crystal.

We evaluate the orbital magnetic moment, $m_{orb}(Co)$, and the spin magnetic moment, $m_{spin}(Co)$, of the $Co^{2+}$ ion using the XMCD sum rules[16,17]:

$$m_{orb} = -\frac{4}{3} \frac{\Delta A_{L_3} + \Delta A_{L_2}}{A_{L_3} + A_{L_2}} \cdot n_h \mu_B, \qquad (1)$$

$$m_{\text{spin}} + 7m_T = -\frac{2\left[\Delta A_{L_3} - 2\Delta A_{L_2}\right]}{A_{L_3} + A_{L_2}} \cdot n_h \mu_B, \tag{2}$$

where $A_{L2}$ ($A_{L3}$), and $\Delta A_{L2}$ ($\Delta A_{L3}$) are the $L_2$- ($L_3$-) edge integrated XAS and XMCD intensities, respectively, $n_h$ is the 3$d$ hole number, and $m_T$ is the magnetic quadrapole moment. By using the integrals[18] $p$, $q$, and $r$ in Fig. 1, and assuming a nominal value of $n_h$ = 3.0 in a Co$^{2+}$ ion, the orbital and the effective spin moments were determined to be $m_{\text{orb}}$(Co) = (0.013±0.002) $\mu_B$, $m_{\text{spin}}$(Co) + 7$m_T$(Co) = (0.12±0.01) $\mu_B$. We thus obtained the total moment $m_{\text{total}}$(Co) = $m_{\text{spin}}$ + $m_{\text{orb}}$ + 7$m_T$ = (0.13±0.01) $\mu_B$. The value of $m_{\text{total}}$ = 0.13 $\mu_B$ is smaller by a factor of ~7-8 than $m_{\text{total}}$(Co) = 1.0 $\mu_B$, determined by magnetization measurements[8]. The origin of this discrepancy is not clear at present, but a possible cause is the formation of a magnetically dead layer on the film surface.

In conclusion, we have verified that the ferromagnetism in a rutile-type Ti$_{0.97}$Co$_{0.03}$O$_{2-\delta}$ film is caused by high-spin Co$^{2+}$ ions substituting the Ti$^{4+}$ ions on the basis of an element-specific experimental and theoretical XMCD study. The present result provides key information for understanding of DMSs based on wide band-gap oxides, and to the development of spintronic devices using them.

This work was done under the approval of the Photon Factory Program Advisory Committee (Proposal No. 2004G206). This work was partially supported by a Grant-in-Aid for Scientific Research (14076209) from the Ministry of Education, Culture, Sports, Science and Technology of Japan.

**Figure captions**

FIG. 1: Co $L_{2,3}$-edge XAS and XMCD of rutile-type $Ti_{0.97}Co_{0.03}O_{2-\delta}$. (a) Photon-helicity ($h$) dependent XAS spectra recorded at 300 K and $B = \pm 1$ T. $\mu_+$ and $\mu_-$ stand for the absorption coefficients with $h$ parallel (↑↑) and antiparallel (↑↓) to the Co $3d$ majority-spin direction. An energy integral $r$ gives $A_{L3} + A_{L2}$, defined in the text. Arrows A through G denote multiplet features. (b) XMCD spectrum and its energy integral. Integrals $p$ and $q$ represent $\Delta A_{L3}$ and $\Delta A_{L3} + \Delta A_{L2}$, respectively, in the text.

FIG. 2: Enlarged plots of spectra in the Co $L_3$ region of $Ti_{0.97}Co_{0.03}O_{2-\delta}$. (a) Co $L_3$-edge XAS. (b) Co $L_3$-edge XMCD. Note the multiplet features denoted by arrows in the XMCD spectrum, corresponding to features A through E in the XAS spectra. This peculiar XMCD spectrum is in sharp contrast to the smooth and featureless XMCD spectrum of metallic Co, which is plotted for a comparison on a scale reduced by a factor of 0.06.

FIG. 3: Comparison of the experimental spectra with atomic multiplet calculations. (a) XAS spectra. (b) XMCD spectra. Calculations were done for the low-spin $Co^{2+}$ ion and the high-spin $Co^{2+}$ ion in crystal fields with $O_h$ symmetry, and the high-spin $Co^{2+}$ ion in a crystal field with $D_{2h}$ symmetry. The calculated XMCD spectra were scaled by a factor of 1/30 for a comparison with the experimental XMCD spectrum.

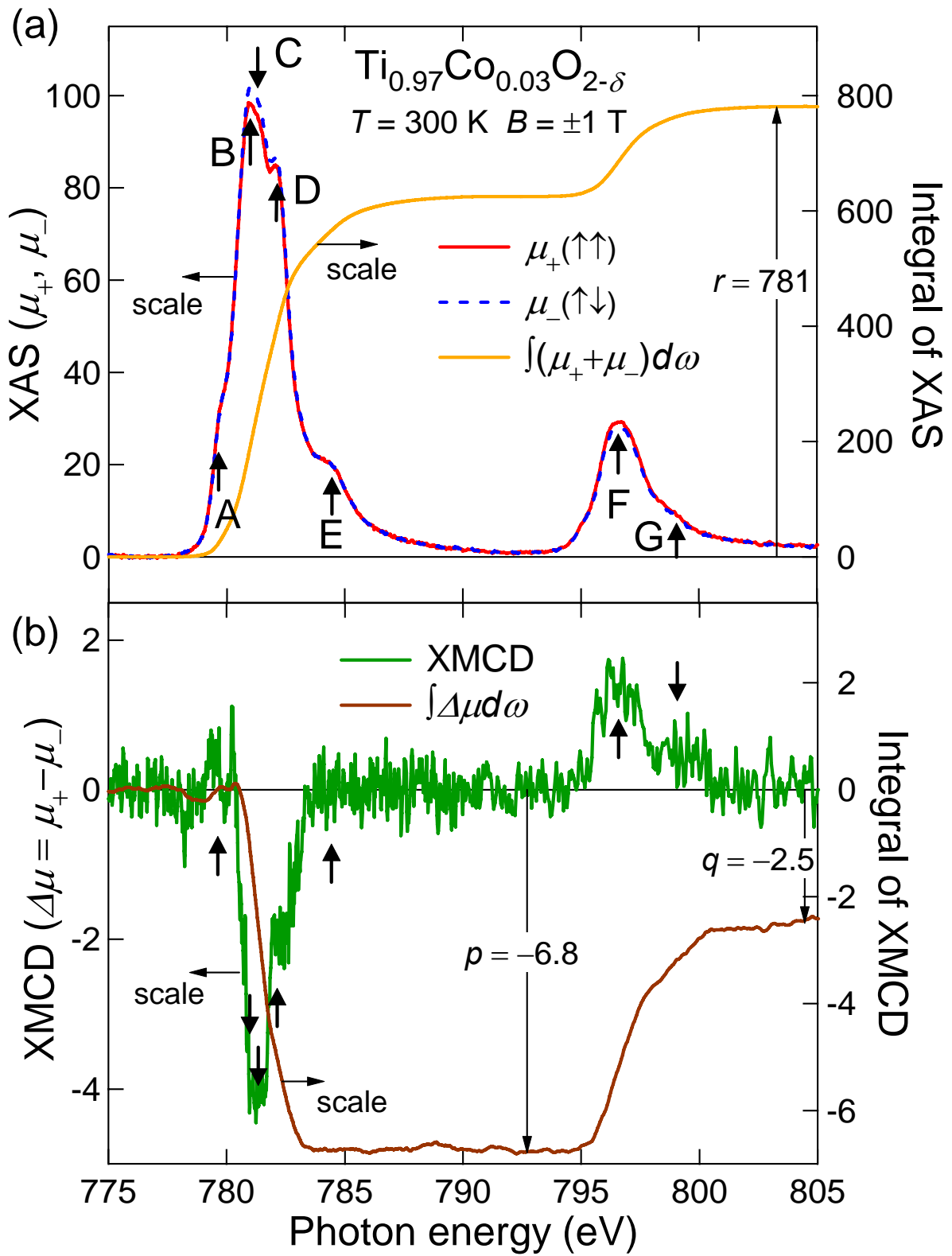

Figure 1 K. Mamiya *et al*.

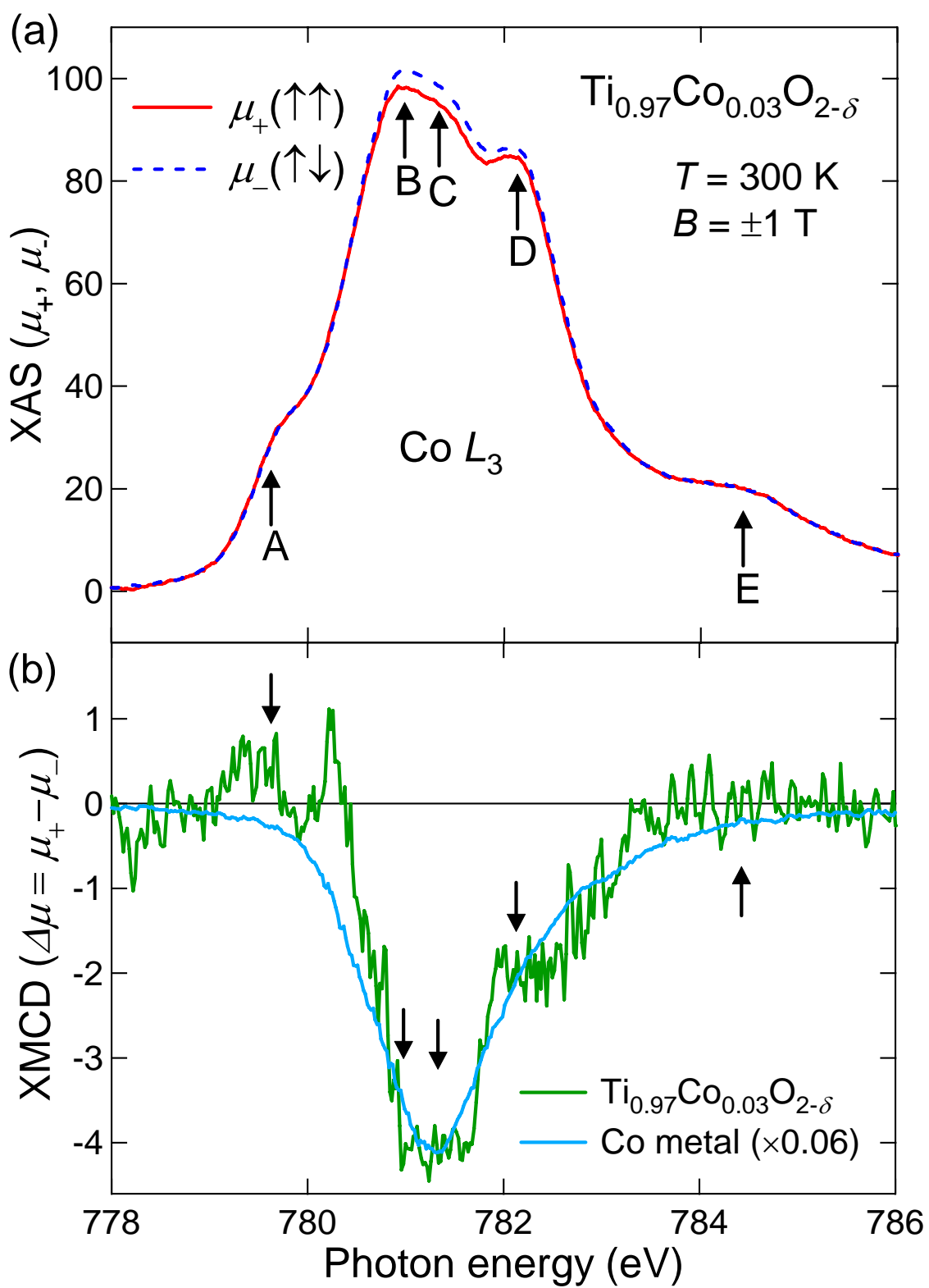

Figure 2 K. Mamiya *et al*.

- 10 -

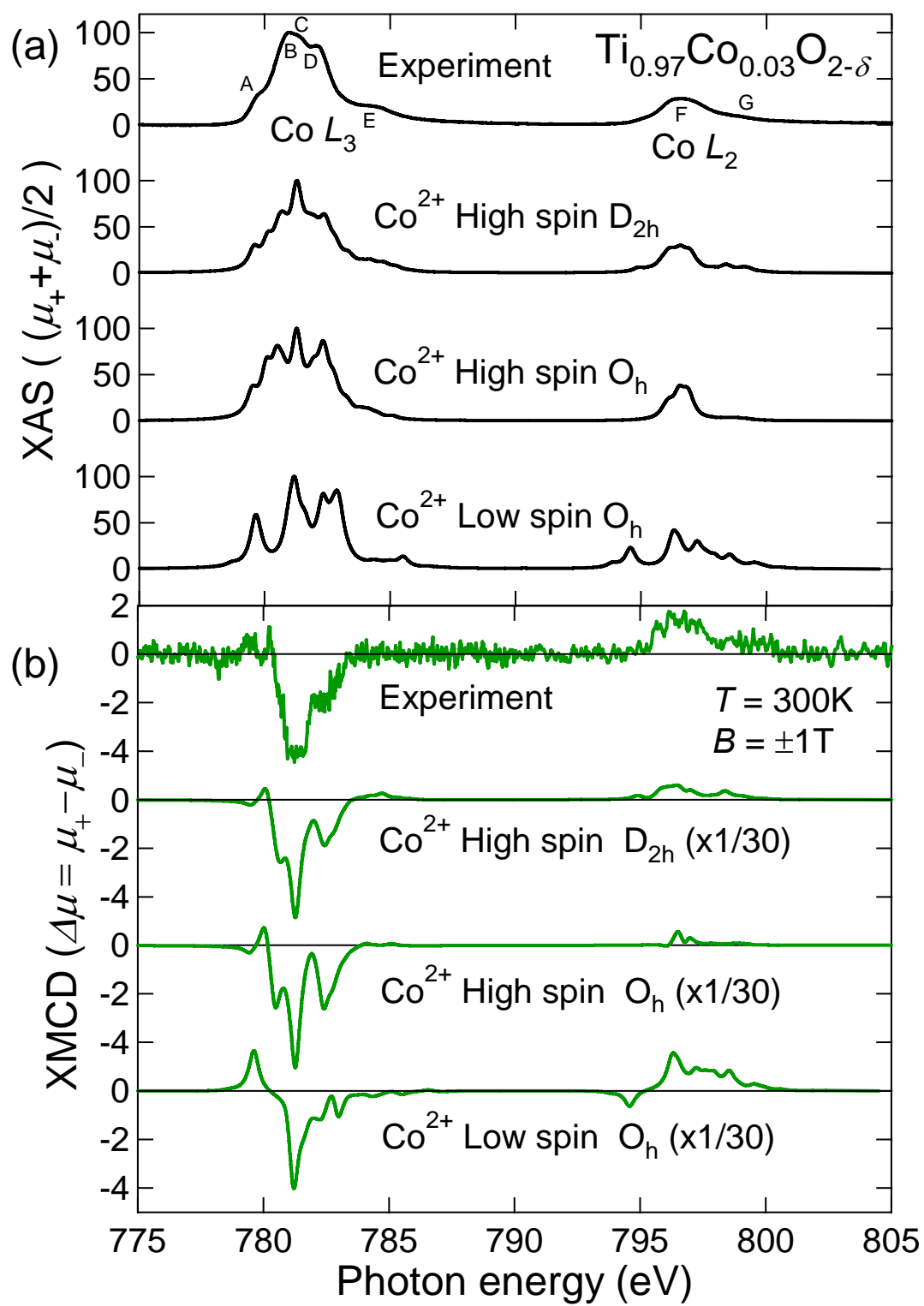

Figure 3 K. Mamiya *et al*.